# Miscibility gap and possible intrinsic Griffiths phase in Sr(Fe$_{1-x}$Mn$_x$)$_2$As$_2$ crystals grown by transitional metal arsenide flux

Long Chen[1,2], Cheng Cao[1,2], Hongxiang Chen[1], Jiangang Guo[1,3], Jie Ma[5], Jiangping Hu[1,3] and Gang Wang[1,3,4,*]

[1]*Beijing National Laboratory for Condensed Matter Physics, Institute of Physics, Chinese Academy of Sciences, Beijing 100190, China*
[2]*University of Chinese Academy of Sciences, Beijing 100049, China*
[3]*Songshan Lake Materials Laboratory, Dongguan, Guangdong 523808, China*
[4]*School of Physical Sciences, University of Chinese Academy of Sciences, Beijing 100049, China*
[5]*Key Laboratory of Artificial Structures and Quantum Control, School of Physics and Astronomy, Shanghai Jiao Tong University, Shanghai 200240, China*

**Abstract:** The crystal structure, magnetic, electronic, and thermal properties of Mn-doped SrFe$_2$As$_2$ crystals have been systematically investigated. A miscibility gap is found in the system from x = 0.4362(4) to x = 0.9612(9). For x < 0.2055(2), the single crystals holding tetragonal structure (space group *I*4/*mmm* (no. 139)) have a continuously enlarged lattice parameter c, followed by a phase separation with crystals holding both tetragonal and trigonal structures up to x = 0.4362(4). Beyond the miscibility gap, the crystals with x > 0.9612(9) hold the trigonal structure (space group *P*-3*m*1 (no. 164)). Moreover, the spin density wave ordering for x = 0 to x = 0.0973(1) is suppressed, followed by an abnormal and broadened increase of the ordering temperature for x = 0.0973(1) to x = 0.2055(2). Eliminating the possibility of real-space phase separation of Mn and Fe based on the results of X-ray diffraction, this novel phenomenon can be attributed to a possible intrinsic Griffiths phase. No any superconducting signals are observed down to 2 K in the whole composition range with 0 < x < 1. A phase diagram with multi-critical points of Mn-doped SrFe$_2$As$_2$ system is established accordingly.

**Introduction**

The discovery of superconductivity [1] and the subsequent enhancement of superconducting (SC) transition temperature ($T_c$) up to 55 K in doped REFeAsO (RE = rare earth) [2-5] have attracted great attention in the research of layered FeAs compounds over the past decade [6-8]. As another FeAs-layered compound with $ThCr_2Si_2$-type (122-type) structure, $AFe_2As_2$ (A = Eu, Sr, Ba, and Ca) [9-13] hosts many similar features including density of states (DOS) near the Fermi surface as those of REFeAsO, which may be attributed to the almost identical FeAs layers [7]. The FeAs layer, serving as a carrier conduction layer [14-17], confines the conduction electrons within quasi-two dimension and leads to strong interactions among these conduction electrons. For the purpose of structure and property tuning of $AFe_2As_2$, substituting A with alkali metals such as K [18-20] or Na [21] is an effective route. Another route is to substitute Fe with transition metals [22]. For instance, electron doping by Co [23,24], Ni [25,26], Rh [27], or Pd [28] at the Fe site in $AFe_2As_2$ (A = Sr, Ba) induces superconductivity, while hole doping by Cr [29,30] or Mn [31,32] for Fe does not, which is the so-called electron-hole asymmetry[33]. In fact, by suppressing the spin density wave (SDW) ordering via doping or pressure [34,35], not only can superconductivity be induced with a dome-like shape $T_c$, but also novel phases will emerge, like the quantum critical point (QCP) [36,37] , the coexisting SC with magnetic states [21,38] or nematic phases [39].

It was reported that the Mn-doped $BaFe_2As_2$ is not SC due to the local Mn moments [40,41], and recent resonant inelastic X-ray scattering (RIXS) studies attributed the lack of superconductivity to the absence of appropriate magnetic excitations [42]. There is a large miscibility gap in $Ba(Fe_{1-x}Mn_x)_2As_2$ with $0.12 \leq x \leq 1$ when slowly cooled down to room temperature (RT) from 1273 K, and the gap narrows to $0.2 \leq x \leq 0.8$ by quenching from 1273 K to 77 K [43]. With increasing Mn, the drop in resistivity of $Ba(Fe_{1-x}Mn_x)_2As_2$ closely related to SDW ordering first switches to a broadened increase, then becomes extremely broad around x = 0.1 and is no longer detectable up to x = 0.147 [44]. For x > 0.1, an evident broad minimum is observed in the derivative of the resistance, and the corresponding transition temperature gradually increases with increasing Mn content [44]. Neutron scattering and high-resolution single-crystalline X-ray diffraction measurements on Ba(Fe$_{1-x}$Mn$_x$)$_2$As$_2$ show the missing tetragonal-to-orthorhombic structural transition whereas the magnetic ordering with a propagation vector of (1/2 1/2 1) persists beyond x = 0.102 [31,41]. Moreover, these unique behaviors can be described by a Griffiths-type phase based on nuclear magnetic resonance (NMR), neutron Larmor diffraction, muon spin resonance (μSR), and inelastic neutron scattering (INS) [45]. Usually, the Griffiths region represents the coexisting disordered paramagnetic region and locally ordered clusters in strongly disordered system. It can alter the critical scaling behavior of a phase transition, or even lead to the appearance of qualitatively new electronic or magnetic states [46-50]. Specially for Ba(Fe$_{1-x}$Mn$_x$)$_2$As$_2$, the so-called Griffiths region corresponds to the coexisting magnetically ordered cluster glass (CG-phase) or spin glass (SG-phase) with paramagnetic regions in the nanoscale. The INS measurement on Ba(Fe$_{1-x}$Mn$_x$)$_2$As$_2$ single crystals with x = 0.075 shows the coexistence of spin excitations at antiferromagnetic (AFM) wave vectors ( $\mathbf{Q}_{AF} = \mathbf{Q}_{stripe} = (\pi, 0)$ ) for BaFe$_2$As$_2$ and ($\mathbf{Q} = \mathbf{Q}_{Neel} = (\pi, \pi)$, rotated 45° from $\mathbf{Q}_{AF}$) for BaMn$_2$As$_2$ [41,51]. Theoretically, a real-space five-band model was built successfully to explain these unexpected properties through a cooperative behavior of the magnetic impurities and the conduction electrons [52].

However, due to the high Neel temperature ( $T_N$ = 625 K ) of BaMn$_2$As$_2$ [53], the effects of G-type AFM order above RT cannot be excluded by neutron diffraction [31,41]. And the similarity between the crystal structures of BaFe$_2$As$_2$ and BaMn$_2$As$_2$ poses a challenge to determine whether this is an intrinsic effect of the system or there is a real-space phase separation between Mn and Fe [51]. Unlike the case with BaFe$_2$As$_2$ (*I*4/*mmm*) and BaMn$_2$As$_2$ (*I*4/*mmm*), the real-space phase separation between Mn and Fe in Sr(Fe$_{1-x}$Mn$_x$)$_2$As$_2$ can be easily determined for the quite different crystal structures of SrFe$_2$As$_2$ (*I*4/*mmm*) and SrMn$_2$As$_2$ (*P*-3*m*1) [54]. And the low G-type AFM ordering temperature ($T_N$ = 125 K) of SrMn$_2$As$_2$ [54] makes it easier to recognize its G-type AFM ordering by neutron diffraction below RT. Mn-doped SrFe$_2$As$_2$ thus potentially provides a better platform to investigate the Griffiths-type phase. Polycrystalline Sr(Fe$_{1-x}$Mn$_x$)$_2$As$_2$ has been studied and no superconductivity is induced even for x higher than 0.20 [55]. Sn-flux-grown Sr(Fe$_{1-x}$Mn$_x$)$_2$As$_2$ single crystals also show no SC signals with suppressed SDW ordering [32]. Given the grain boundary effects of polycrystalline sample and the probable Sn contamination of Sn-flux-grown sample [56-58], we decided to systematically investigate the structural, magnetic, electronic, and thermal properties of Sr(Fe$_{1-x}$Mn$_x$)$_2$As$_2$ crystals grown by transition metal arsenide (TMA) flux and carefully study the possible Griffiths-type phase or the possible superconductivity. We observed the suppressed SDW order with ordering temperature reduced to about 140 K at x = 0.0973(1). For 0.0973(1) ≤ x < 0.2055(2), the following increasing ordering temperature with broaden feature is attributed to the Griffiths phase like that of Ba(Fe$_{1-x}$Mn$_x$)$_2$As$_2$. The real-space phase separation between Mn and Fe was observed only for 0.2055(2) ≤ x ≤ 0.4362(4), hence this Griffiths phase is probably being an intrinsic property of Sr(Fe$_{1-x}$Mn$_x$)$_2$As$_2$ system. A miscibility gap ranging from x = 0.4362(4) to x = 0.9612(9) was observed and no any SC signals were found down to 2 K in the whole composition range with 0 ≤ x ≤ 1. Finally, a global phase diagram of TMA-flux-grown Sr(Fe$_{1-x}$Mn$_x$)$_2$As$_2$ crystals with multi-critical points was established.

**Experiment**

Mn-doped SrFe$_2$As$_2$ crystals were grown by a high-temperature solution method using the TMA flux. The strontium pieces (Alfa Aesar, 99%), and ground, preheated FeAs and MnAs precursors were mixed by a molar ratio of 1 : 4(1-x) : 4x in a fritted alumina crucible set (Canfield Crucible Set or CCS) [59] and sealed in a fused-silica ampoule at vacuum. FeAs or MnAs precursor was prepared by heating the mixture of arsenic powder (Alfa Aesar, 99.9999+%) and ion powder (Alfa Aesar, 99.9+%) or manganese powder (Alfa Aesar, 99.95%) at 1173 K and 1073 K for 10 hours, respectively. The sealed ampoule was heated to 1423 K, kept for 5 hours, and then slowly cooled down to 1273 K at a rate of 5 K/h. By centrifugation at 1273 K, black, shiny, and air-stable plate-like crystals as large as 4 mm × 4 mm × 0.5 mm were obtained.

Single crystal X-ray diffraction (SCXRD) patterns were collected using a Bruker D8 VENTURE diffractometer with multilayer monochromatized Mo K$_\alpha$ ($\lambda$ = 0.71073 Å) radiation. Unit cell refinement and data merging were done with the SAINT program, and an absorption correction was applied by Multi-Scans scanning. Powder X-ray diffraction (PXRD) data were collected on a PANalytical X'Pert PRO diffractometer (Cu K$_\alpha$ radiation) operated at 40 kV voltage and 40 mA current with a graphite monochromator in a reflection mode (2$\theta$ = 5 – 100º, step 0.017º) [60]. Indexing and Rietveld refinement were performed using the DICVOL91 and FULLPROF programs [61]. The elemental analysis was carried out using a scanning electron microscope (SEM, Hitachi S-4800) equipped with an electron microprobe analyzer for the semiquantitative elemental analysis in the energy-dispersive spectroscopy (EDS) mode and inductively coupled plasma-atomic emission spectrometer (ICP-AES, Teledyne Leeman Labs Prodigy 7). For each crystal with certain doping level, 5 spots in different areas were measured by EDS. The ICP-AES measurement was carried out on two pieces of single crystal at each doping level. Temperature-dependent electronic resistance and heat capacity measurements were carried out on a physical property measurement system (PPMS, Quantum Design), whereas field-dependent magnetization measurements being carried out on a vibrating sample magnetometer system (VSM, Quantum Design). Contacts for standard four-probe configuration were made by attaching platinum wires using silver paint, resulting in a contact resistance smaller than 5 Ω. Samples for all the measurements were cleaved along (00l) from Sr(Fe$_{1-x}$Mn$_x$)$_2$As$_2$ crystals using a razor blade.

**Results and discussion**

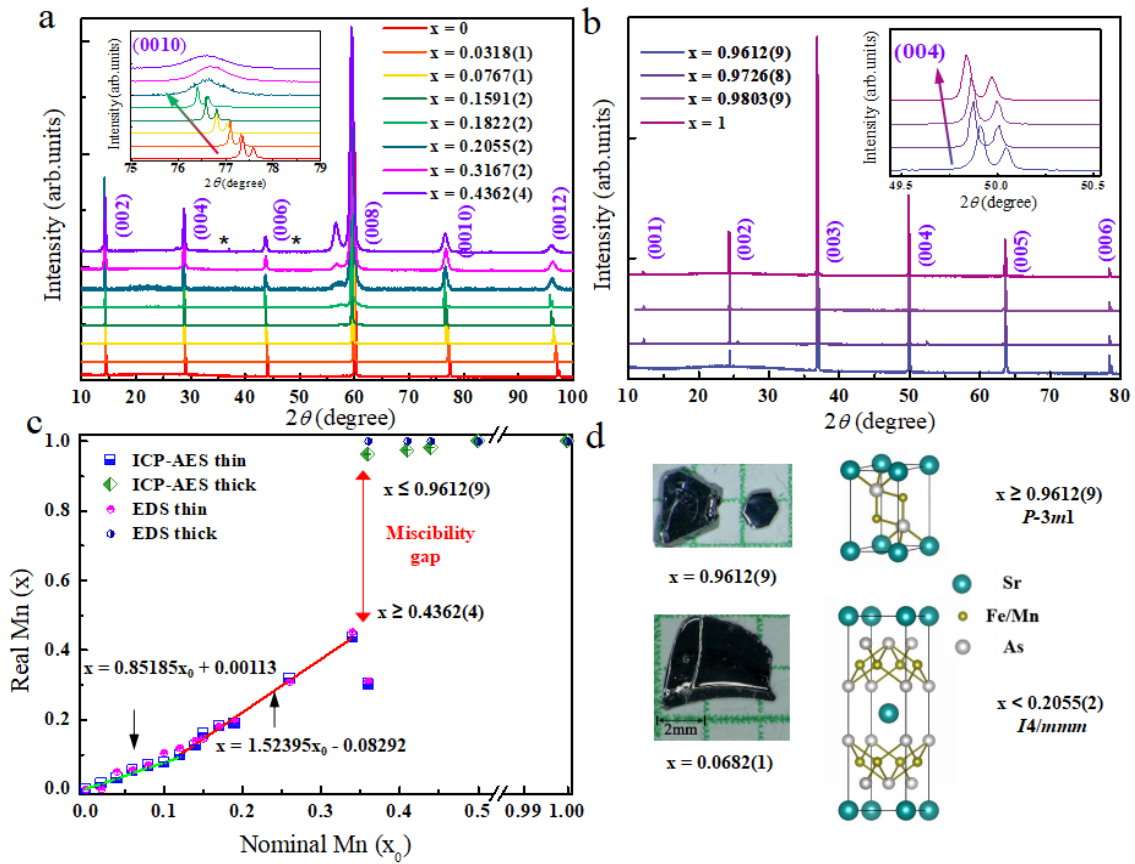

Figure 1. (a) (Color online) The XRD patterns of Sr(Fe$_{1-x}$Mn$_x$)$_2$As$_2$ (Fe-rich side, x ≤ 0.4362(4)) crystals showing (00l) (l = even number) diffraction peaks. The inset is the enlarged (0010) diffraction peak. The star symbols show the (003) and (004) diffraction peaks of Mn-rich phase, indicating the phase separation for x ≥ 0.2055(2). (b) The XRD patterns of Sr(Fe$_{1-x}$Mn$_x$)$_2$As$_2$ (Mn-rich phase, x ≥ 0.9612(9)) crystals showing (00l) (l = integer) reflections. The inset is the enlarged (004) diffraction peak. The spectra in (a) and (b) are offset for clarity. (c) Real Mn content in Sr(Fe$_{1-x}$Mn$_x$)$_2$As$_2$ crystals determined by ICP-AES and EDS. The lines are the linear fittings of ICP-AES results and the double arrow (red) indicates the miscibility gap. (d) The optical photographs of the typical thin and thick crystals with x = 0.0682(1) and x = 0.9612(9) and corresponding schematic crystal structures for x < 0.2055(2) and x ≥ 0.9612(9).

Figure 1 shows the crystal structure and chemical composition of Sr(Fe$_{1-x}$Mn$_x$)$_2$As$_2$ crystals. The XRD patterns of Sr(Fe$_{1-x}$Mn$_x$)$_2$As$_2$ crystals for x ≤ 0.4362(4) and x ≥ 0.9612(9) are shown in Figure 1(a) and (b), respectively. These two different series patterns can be indexed on a tetragonal structure (*I*4/*mmm*) and a rhombohedral structure (*P*-3*m*1) respectively, as schematically displayed in Figure 1(d). The (00l) diffraction peaks with even l (x ≤ 0.4362(4)) or integer l (x ≥ 0.9612(9)) are observed, indicating the plate surface is perpendicular to the c axis. These peaks are in good agreement with those of $SrFe_2As_2$ [10] and $SrMn_2As_2$ [62] polycrystalline samples as reported. As shown by the enlarged (0010) diffraction peak, the peak position gradually shifts to a lower angle with increasing Mn content (x < 0.2055(2)), indicating that the lattice parameter *c* expands. Then the diffraction peaks broaden at higher doping levels (0.2055(2) ≤ x ≤ 0.4362(4)) with enlarged shoulder for (008) diffraction peak (Figure 1(a)). The appearance of the (003) and (004) diffraction peaks of Mn-rich $Sr(Fe_{1-x}Mn_x)_2As_2$ indicates the phase separation in the corresponding composition. At the Mn-rich side (x ≥ 0.9612(9)), the (004) diffraction peak also shows low-angle shift with increasing Mn content (inset of Figure 1(b)).

Figure 1(c) and Table S1 show the real Mn content of $Sr(Fe_{1-x}Mn_x)_2As_2$ crystals determined by ICP-AES and EDS. The contents of Mn (x) determined by these two techniques are consistent with each other and have good linearity in each region. For ICP-AES results, it is found that the real Mn content (x) is lower than the nominal one ($x_0$) before x = 0.1254(1), and higher than the nominal one for 0.1591(2) ≤ x ≤ 0.4362(4). No crystals with 0.4362(4) < x < 0.9612(9) are obtained, indicating the existence of a large miscibility gap but smaller than that of $Ba(Fe_{1-x}Mn_x)_2As_2$ [43]. Details of phase separation and miscibility gap can be seen in Supporting Information Figure S1. Beyond the miscibility gap, there is a small composition range for Fe-doped $SrMn_2As_2$. We then summarized the structural transition for the $Sr(Fe_{1-x}Mn_x)_2As_2$ system accordingly. With increasing Mn content, $Sr(Fe_{1-x}Mn_x)_2As_2$ evolves from *I*4/*mmm* (no. 139) isostructural to $SrFe_2As_2$, to a two-phase-coexisted region and a miscibility gap, and finally to *P*-3*m*1(no. 164) isostructural to $SrMn_2As_2$.

Figure 1(d) shows the optical photographs of the typical thin and thick crystals with corresponding x = 0.0682(1) and 0.9612(9). The thin crystal can be as large as 4 mm × 4 mm × 0.5 mm, whereas the thick crystal exhibits hexagonal appearance. Using

SCXRD, the thin crystals with x < 0.2055(2) are indexed with the space group *I*4/*mmm* (no. 164), whereas the thick crystals with x ≥ 0.9612(9) characterized by the *P*-3*m*1 (no. 139) space group. (See Supporting Information Table SII.)

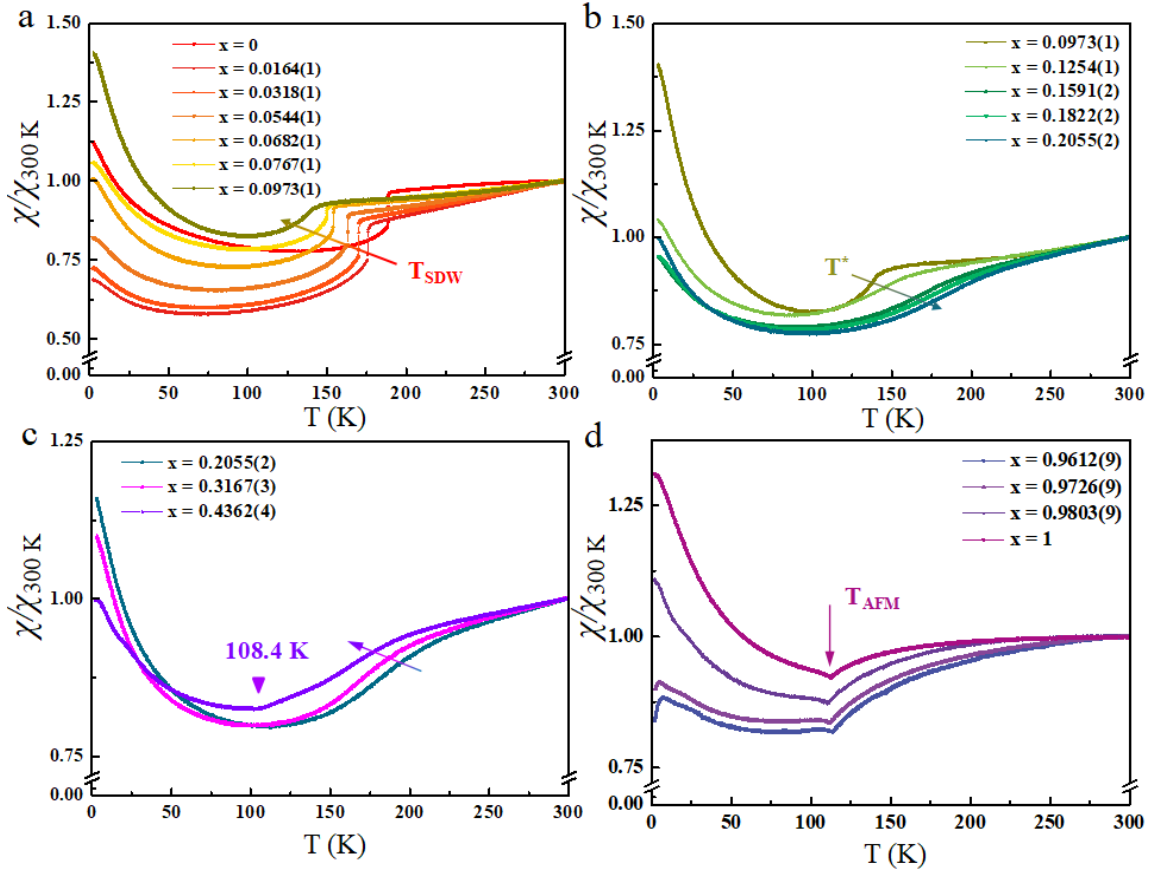

Figure 2. (Color online) Normalized temperature-dependent dc magnetic susceptibility ($\chi/\chi_{300K}$) of Sr(Fe$_{1-x}$Mn$_x$)$_2$As$_2$ crystals with an applied magnetic field of 7 T parallel to ab plane in the ZFC protocol. (a) 0 ≤ x ≤ 0.0973(1), (b) 0.0973(1) ≤ x ≤ 0.2055(2), (c) 0.2055(2) ≤ x ≤ 0.4362(4), and (d) 0.9612(9) ≤ x ≤ 1.

Figure 2 shows the normalized temperature-dependent dc magnetic susceptibility ($\chi/\chi_{300K}$) of Sr(Fe$_{1-x}$Mn$_x$)$_2$As$_2$ crystals measured under an applied field of 7 T parallel to the ab plane in the zero-field-cooling (ZFC) protocol. The magnetic transition temperatures are determined by maxima in $d(\chi T)/dT$ (See Supporting Information Figure S2. (a) and (b)). No hint of superconductivity is observed in the whole composition range. For 0 ≤ x ≤ 0.0973(1), the drop associated with the coincident SDW order and structural transition is suppressed from 188.6 K for x = 0 to 138.8 K for x = 0.0973(1) (Figure 2(a)). For x = 0.0973(1), the drop has broadened and become difficult to resolve in $\chi/\chi_{300K}$ curve but is still clear in $d(\chi T)/dT$. Following this, a new broadened magnetic feature with increasing transition temperature emerges with increasing Mn content (x) from x = 0.0973(1) to x = 0.2055(2) (Figure 2(b) and Figure S2. (b)). Then the magnetic transition shows a broaden decrease trend at 0.2055(2) ≤ x ≤ 0.4362(4) accompanied by phase separation (Figure 2(c) and Figure 1(a)). For x = 0.4362(4), there is an anomaly at 108.4 K, which can be attributed to the AFM ordering of Mn-rich phase due to the real-space separation of Mn and Fe. In the miscibility gap, the crystals in one batch show different magnetism (See Supporting Information Figure S3 (a) and (b)). When x is beyond the miscibility gap, the Mn-rich crystals exhibit an magnetic transition (Figure 2(d)) around 110 K, which corresponds to the reported antiferromagnetism of SrMn$_2$As$_2$ [54]. The magnetic feature at low temperature may be related to FeAs impurity in the remaining flux droplet.

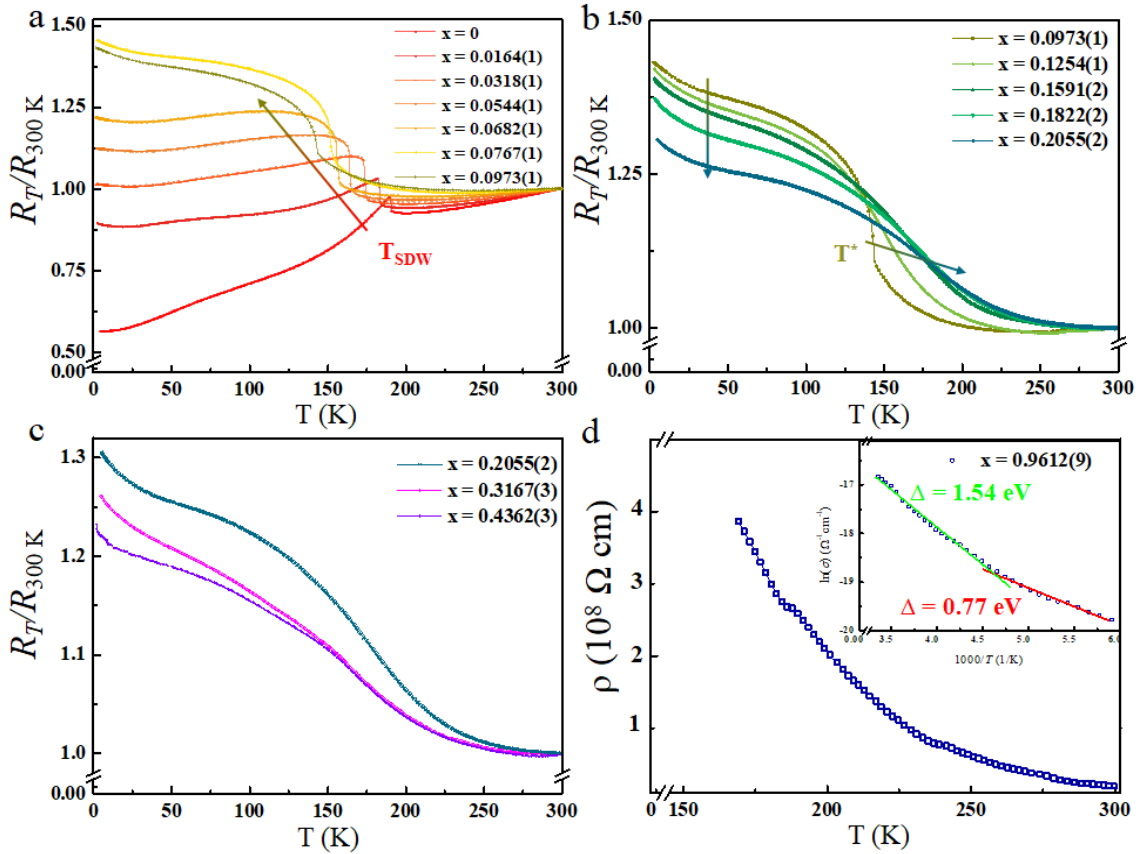

Figure 3. (Color online) Normalized temperature-dependent in-plane resistance ($R_T/R_{300\,K}$) of Sr(Fe$_{1-x}$Mn$_x$)$_2$As$_2$ crystals. (a) 0 ≤ x ≤ 0.0973(1), (b) 0.0973(1) ≤ x ≤ 0.2055(2), (c) 0.2055(2) ≤ x ≤ 0.4362(4), and (d) x = 0.9612(9). The inset shows ln(σ)

versus inverse temperature 1000/$T$ (where conductivity $\sigma = 1/\rho$); the solid curves are fittings over the temperature interval by the expression $\ln(\sigma) = A - \Delta/T$, where $A$ is a constant and $\Delta$ is the activation energy.

Figure 3 shows the normalized temperature-dependent in-plane resistance ($R_T/R_{300\ K}$) for Sr(Fe$_{1-x}$Mn$_x$)$_2$As$_2$ crystals. The coupled SDW ordering and structure transition temperature is determined by minima in $dR/dT$ (See Supporting Information Figure S2. (c) and (d)). The signal of superconductivity is not observed in the whole composition range, which is consistent with the magnetization measurements. With Mn doping, the SDW ordering temperature decreases for x ≤ 0.0973(1) with an enhancement in low-temperature resistance (Figure 3(a)). The resistance curves gradually shift from a sharp decrease to a broadened increase. Then a feature with increasing characteristic temperature emerges at x = 0.0973(1), and maintains up to x = 0.2055(2) (Figure 3(b) and Figure S2. (d)), being consistent with the magnetization measurements. No clear transition is observed in $R_T/R_{300\ K}$ for 0.2055(2) < x ⩽ 0.4362(4) (Figure 3(c)), but a decrease trend of the minima in $dR/dT$ can be observed (Figure S2. (d)). Signal of the miscibility gap was also observed in the temperature-dependent resistivity (See Supporting Information Figure S3 (c) and (d)). At $x_0$ = 0.36, crystals with different appearance and composition growing in one batch behave differently in electronic transportation. The resistivity of thin crystal with x = 0.3016(3) gradually increases with the decreasing temperature, whereas the resistivity of the thick crystal with x = 0.9612(9) overflows at about 170 K (Figure 3(d)) and behaves like a semiconductor. By fitting the conductivity $\sigma = 1/\rho$ with the expression $\ln(\sigma) = A - \Delta/T$ in two different temperature ranges, we obtained $\Delta$ = 0.77 eV from 170 K to about 210 K and $\Delta$ = 1.54 eV from 210 K to 300 K. The larger one is deduced to the intrinsic activation energy, and the smaller one being the energy gap between the donor energy levels and conduction band or the acceptor energy levels and valence band [63,64]. Both of them have the same order of magnitude as 0.29 eV (0.65 eV) within the temperature range from 100 K to 285 K [54] and are much larger than 85 meV fitted by the expression $\log_{10}\rho = A + 2.303\Delta/T$ between 70 K and 120 K [65] for Sn-flux-grown SrMn$_2$As$_2$. The difference is most probably due to inevitable deviation in measurements or the flux effect as reported for BaMn$_2$As$_2$ [64].

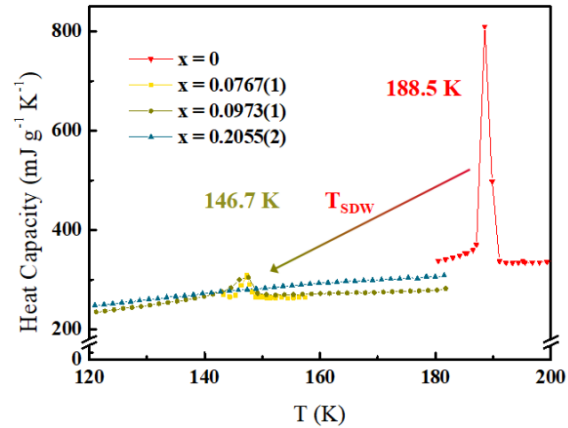

Figure 4. (Color online) Heat capacity of Sr(Fe$_{1-x}$Mn$_x$)$_2$As$_2$ crystals for x ≤ 0.2055(2). The difference in background may be owing to the relatively high measuring temperature and mass weighing deviation.

Figure 4 shows the heat capacity of some Sr(Fe$_{1-x}$Mn$_x$)$_2$As$_2$ crystals with x = 0, 0.0767(1), 0.0973(1), and 0.2055(2). The suppression of SDW ordering temperature is clearly observed for x ≤ 0.0973(1), being consistent with the magnetism and resistance measurements. The only peak shows that the structural transition and SDW ordering are coupled, meanwhile the broadened and weakened tendency of the peak from x = 0 to x = 0.0973(1) indicates the suppression of such transition. At x = 0.2055(2), there is no anomaly observed, which suggests there is no structural transition and long-range magnetic ordering.

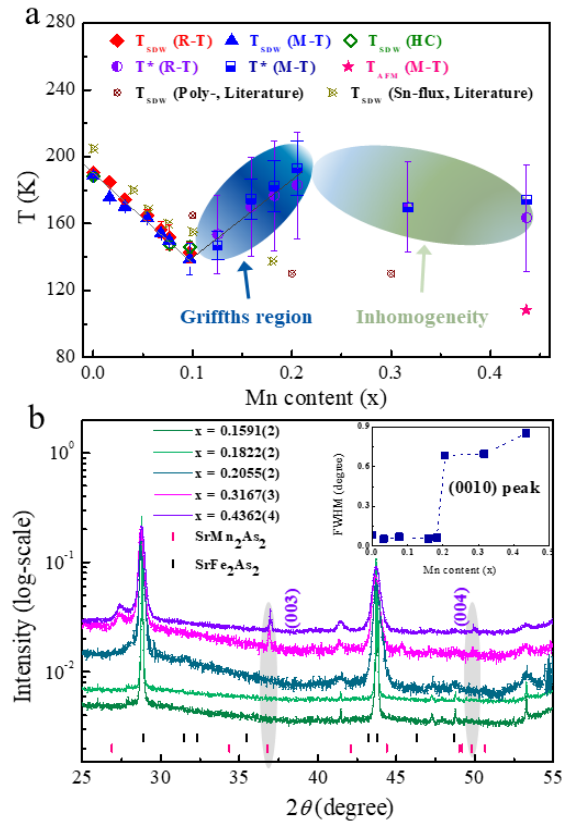

Figure 5. (Color online) (a) The Mn-content-dependent critical temperature (T versus x) before the miscibility gap (x ≤ 0.4362(4)). The transition temperatures for polycrystalline $Sr(Fe_{1-x}Mn_x)_2As_2$ ($T_{SDW}$ (Poly-, Literature)) and Sn-flux-grown $Sr(Fe_{1-x}Mn_x)_2As_2$ single crystals ($T_{SDW}$ (Sn-flux, Literature)) are shown for comparison [32,55]. $T_{SDW}$ donates the SDW ordering temperature, $T^*$ donates the unusually enhanced ordering temperature beyond x = 0.0973(1), and $T_{AFM}$ donates the AFM ordering temperature. The solid lines are the linear fitting of critical temperature at different ranges of Mn content (x). Shaded areas with different colors outline the Griffiths region and phase-separated region for $Sr(Fe_{1-x}Mn_x)_2As_2$. (b) The log-scale XRD patterns of $Sr(Fe_{1-x}Mn_x)_2As_2$ crystals (only 0.1591(2) ≤ x ≤ 0.4362(4) for comparison). The short vertical lines show the peak positions corresponding to $SrFe_2As_2$ and $SrMn_2As_2$ from ICSD Data Base [10,62] and the shaded areas indicate the (003) and (004) peaks for $SrMn_2As_2$. These spectra are offset for clarity. The inset shows the Mn-content-dependent full wavelength of half maximum (FWHM versus x) of (0010) peak.

Based on the magnetism, electronic transportation, and heat capacity measurements,

the Mn-content-dependent critical temperature (T versus x) was summarized (Figure 5 (a)). The linear suppression of SDW ordering temperature to about 140 K before x = 0.0973(1) is well consisted with polycrystalline samples [55] and Sn-flux-grown crystals [32]. Then the critical temperature shows an unusual increase with broaden feature, which has never been reported previously in this system. The suppression of SDW order and the followed unusual increase of critical temperature before and after the critical point x = 0.0973(1) are quite similar to those in Ba(Fe$_{1-x}$Mn$_x$)$_2$As$_2$. For Ba(Fe$_{1-x}$Mn$_x$)$_2$As$_2$, the ordering temperature is suppressed from 134 K down to about 50 K at x = 0.102, followed by an enhancement to about 100 K at x = 0.147 [44], which has been attributed to the Griffiths effect [45,47,48]. As for Sr(Fe$_{1-x}$Mn$_x$)$_2$As$_2$, the ordering temperature is suppressed from 198 K down to about 140 K at x = 0.0973(1), followed by an enhancement to about 175 K at x = 0.2055(2). Considering these similarities between the two systems, we would attribute the unusual increase of ordering temperature in Sr(Fe$_{1-x}$Mn$_x$)$_2$As$_2$ system to the Griffiths region as that of Ba(Fe$_{1-x}$Mn$_x$)$_2$As$_2$.

In Ba(Fe$_{1-x}$Mn$_x$)$_2$As$_2$, it is very difficult to determine the Griffiths effect as an intrinsic effect or a real-space separation between Mn and Fe [51]. Fortunately, for Sr(Fe$_{1-x}$Mn$_x$)$_2$As$_2$, XRD patterns could clearly determine the real-space separation between Mn and Fe due to the different lattice symmetry of the two parent compounds. As shown in Fig. 5 (b), the XRD patterns indicate that the diffraction peaks for samples with x = 0.1591(2) and x = 0.1822(2) are narrow and have nearly unchanged stable full wavelength of half maximum (FWHM), indicating good homogeneity. Whereas the XRD patterns of samples with 0.2055(2) ≤ x ≤ 0.4362(4) present about ten times larger FWHM and the coexistence of tetragonal and rhombohedral phases. The narrow and unchanged FWHM shows good homogeneity for x < 0.2055(2), and the coexisting phases for 0.2055 ≤ x ≤ 0.4362(4) donate the inhomogeneity (Fig. 5 (a)). As pointed out by D. S. Inosov et al. [45], in the homogeneity region, the Griffiths effect will enhance the ordering temperature with broaden feature, whereas inhomogeneity give rise to a decrease trend of ordering temperature.

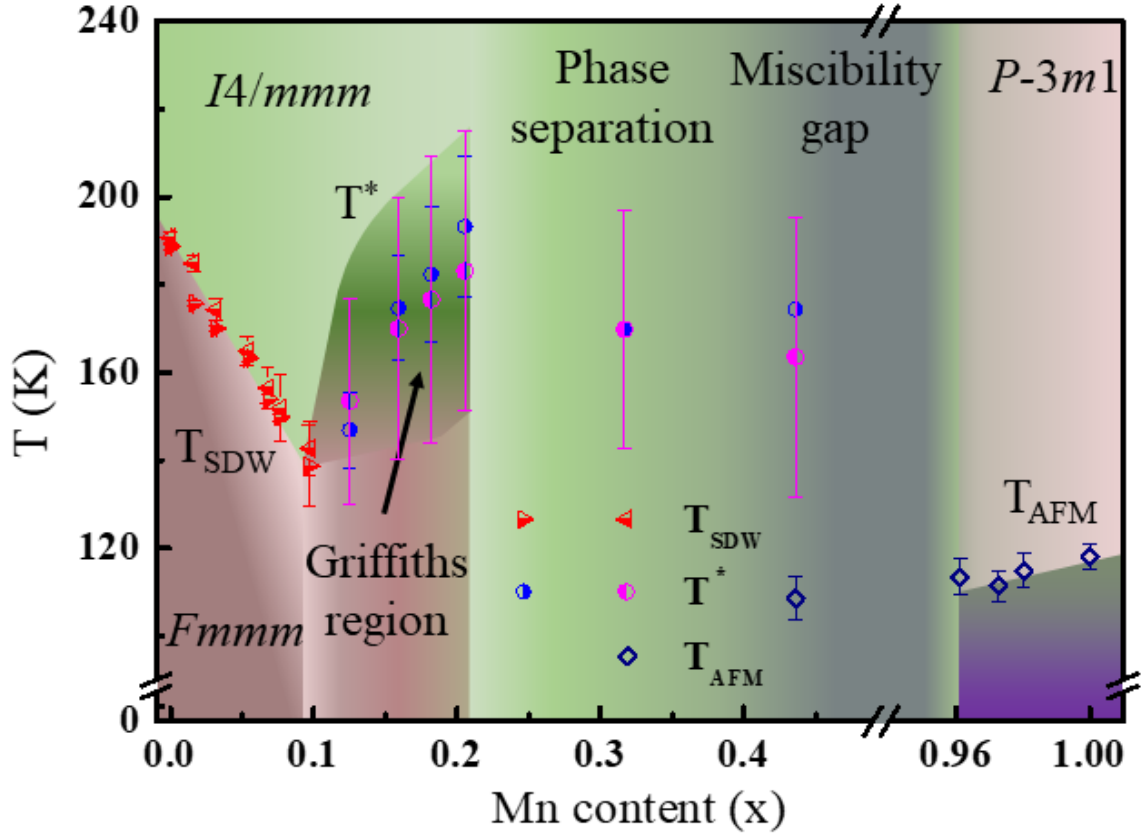

Figure 6. (Color online) The schematic phase diagram of TMA-flux-grown Sr(Fe$_{1-x}$Mn$_x$)$_2$As$_2$ crystals.

Combining the results presented above, the schematic phase diagram of TMA-flux-grown Sr(Fe$_{1-x}$Mn$_x$)$_2$As$_2$ crystals is established. As shown in Figure 6, no superconductivity has been observed in the whole composition range. For $0 \leq x < 0.0973(1)$, Sr(Fe$_{1-x}$Mn$_x$)$_2$As$_2$ has the similar crystal structure and physical properties as SrFe$_2$As$_2$ with suppressed SDW ordering temperature. Meanwhile, the lattice parameter $c$ expands with the increasing x. For $0.0973(1) \leq x \leq 0.2055(2)$, a broadened magnetic feature with increasing transition temperature emerges. This feature could be attributed to the intrinsic Griffiths effect. Unlike Mn-doped BaFe$_2$As$_2$, the possibility of real-space separation between Mn and Fe is eliminated based on the results of XRD. For $0.2055(2) < x \leq 0.4362(4)$, the real-space separation of Mn and Fe is observed. For $0.4362(4) < x < 0.9612(9)$, a miscibility gap does exist, and the grown crystals in one batch have different crystal structures and properties. For $0.9612(9) \leq x \leq 1$, Sr(Fe$_{1-x}$Mn$_x$)$_2$As$_2$ crystals have the similar structure to that of SrMn$_2$As$_2$ with robust AFM order and semiconductor-like transport behavior.

For Mn-doped BaFe$_2$As$_2$, the enhanced ordering temperature (T$^*$) above a critical Mn concentration, as well as other properties like the missing tetragonal-to-orthorhombic structural transition beyond the critical concentration and the unexpected G-type spin excitation (Q$_{Neel}$) at low concentration are successfully explained by a real-space five-band model [52] building by combing the Ruderman-Kittel-Kasuya-Yosida (RKKY) exchange interactions in multi orbital nested systems at the brink of an instability [66,67] with the physics of AFM Griffiths effect in itinerant systems [52,68]. Here we observed the similar unusual enhancement of ordering temperature (T$^*$) beyond a critical doping concentration caused by Griffiths effect. More investigations, like the Synchrotron diffraction, Raman scattering, or INS will be needed to further understand the origin of the Griffiths phase. Furthermore, it is predicted that the Griffiths effect can smear the nematic quantum phase transition in inhomogeneous system [69] and Sr(Fe$_{1-x}$Mn$_x$)$_2$As$_2$ system could be a candidate to fulfill that. Compared with other transition metal doped AFe$_2$As$_2$ (A = Eu, Sr, Ba, and Ca) systems [25-27,55,57,70-73], the behavior caused by Griffiths effect has only been found in Mn-doped systems [45], to the best of our knowledge. We believe that the results above will be helpful to understand the uniqueness of Mn-doped systems and figure out the origin of Griffiths phase, or to explore more novel phases.

Beyond AFe$_2$As$_2$ (A = Eu, Sr, Ba, and Ca) system, it is found that tiny amount of Mn (0.2%) points toward the presence of a QCP at the crossover between SC and magnetic states in LaFe$_{1-x}$Mn$_x$AsO$_{0.89}$F$_{0.11}$ [36]. Mn doped K$_{0.8}$Fe$_2$Se$_2$ will lead to the magnetic pairing breaking, and T$_c$ being suppressed rapidly [74,75]. By a two-step substitution in (Li, Fe)OHFeSe, first the tetrahedral site in (Li,Fe)OH layers and then the site in the SC-dominating FeSe layers, lightly Mn-substituted (Li, Fe)OHFeSe single crystals show a V-shaped change in the lattice parameter and a re-enhancement of T$_c$ for 0.02 < x < 0.07 [76]. For Sr(Fe$_{1-x}$Mn$_x$)$_2$As$_2$, there is no signal of superconductivity observed in the whole composition range, which may be attributed to the non-SC parent phase SrFe$_2$As$_2$. Therefore, high pressure may be useful to realize the QCP or superconductivity in this doped system.

**Conclusion**

In summary, the crystal structure, magnetic, electronic, and thermal properties of Sr(Fe$_{1-x}$Mn$_x$)$_2$As$_2$ crystals grown by TMA flux have been systematically investigated. No signal of superconductivity is observed at temperature down to 2 K. The substitution of Fe by Mn first suppresses the SDW order before a critical point at x = 0.0973(1). Then, a broadened ordering temperature appears, which implies the intrinsic Griffiths region for the lack of real-space separation of Mn and Fe. With higher Mn content (x ≥ 0.2055(2)), the real-space separation of Mn and Fe is observed for the co-existing tetragonal and rhombohedral structures. In the following, a large miscibility gap from x = 0.4362(4) to x = 0.9612(9) as that in Mn-doped BaFe$_2$As$_2$ emerges. Beyond the miscibility gap, the Fe-doped SrMn$_2$As$_2$ exhibits a robust AFM order. Accordingly, a global phase diagram of TMA-flux-grown Sr(Fe$_{1-x}$Mn$_x$)$_2$As$_2$ with multi-critical points is established. Sr(Fe$_{1-x}$Mn$_x$)$_2$As$_2$ provides a platform to investigate the origin of the Griffiths-type phase and to fulfill novel phases like the smeared nematic quantum phases or superconductivity.

**Acknowledgement**

L. Chen and G. Wang acknowledge Prof. X. L. Chen of the Institute of Physics, CAS for discussion. This work was partially supported by the National Natural Science Foundation of China (Grant Nos. 51572291 and 51832010), the National Key Research and Development Program of China (Grant No. 2017YFA0302902 and 2018YFE0202600), and the Key Research Program of Frontier Sciences, Chinese Academy of Sciences (Grant No. QYZDJ-SSW-SLH013).

gangwang@aphy.iphy.ac.cn

# Miscibility gap and possible intrinsic Griffiths phase in Sr(Fe$_{1-x}$Mn$_x$)$_2$As$_2$ crystals grown by transitional metal arsenide flux

Long Chen, Cheng Cao, Hongxiang Chen, Jiangang Guo, Jie Ma, Jiangping Hu, and Gang Wang

**Supporting Information**

Table SI. Real Mn content (x) of $Sr(Fe_{1-x}Mn_x)_2As_2$ crystals determined by ICP-AES and EDS. The ICP-AES analysis gives the mean of two crystals with the same nominal Mn content ($x_0$), σ being the error decided by accuracy of ICP-AES. The EDS analysis gives the mean of five measured values on one crystal at different area, σ being the standard deviation of these 5 measured values. The accuracy of EDS is about 5%, so real Mn content (x) lower than 5% is not determinable by EDS.

| Nominal Mn ($x_0$) | Mn (ICP-AES, x) | | Mn (EDS, x) | |
|---|---|---|---|---|
| | thin | thick | thin | thick |
| 0 | 0 | | 0 | |
| 0.02 | 0.0164(1) | | 0(0) | |
| 0.04 | 0.0318(1) | | 0.050(1) | |
| 0.06 | 0.0544(1) | | 0.054(1) | |
| 0.08 | 0.0682(1) | | 0.069(1) | |
| 0.1 | 0.0767(1) | | 0.104(1) | |
| 0.12 | 0.0973(1) | | 0.118(1) | |
| 0.14 | 0.1254(1) | | 0.139(1) | |
| 0.15 | 0.1591(2) | | 0.145(1) | |
| 0.17 | 0.1822(2) | | 0.181(2) | |
| 0.20 | 0.2055(2) | | 0.198(2) | |
| 0.26 | 0.3167(3) | | 0.309(3) | |
| 0.34 | 0.4362(4) | | 0.452(5) | |
| 0.36 | 0.3016(3) | 0.9612(9) | 0.306(3) | 1.0 |
| 0.41 | | 0.9726(9) | | 1.0 |
| 0.44 | | 0.9803(9) | | 1.0 |
| 0.50 | | 1.0 | | 1.0 |
| 1.00 | | 1.0 | | 1.0 |

Table SII. Crystal structures of Sr(Fe$_{1-x}$Mn$_x$)$_2$As$_2$ (x = 0.0682(1), 0.0973(1), and 1) obtained by structural refinement. The solution and refinement of the crystal structure were carried out using the SHELX suite [1] and Jana2006 [2]. The final refinement was performed using anisotropic atomic displacement parameters for all atoms. A summary of pertinent information relating to the formula, unit cell parameters, data collection, and refinements is provided in Table SII.

| Sr(Fe$_{1-x}$Mn$_x$)$_2$As$_2$ | x = 0.0682(1) | x = 0.0973(1) | x = 1 |
|---|---|---|---|
| Formula weight | 349.0262 | 348.9899 | 347.3393 |
| Space Group No. | *I*4/*mmm* (139) | *I*4/*mmm* (139) | *P*-3*m*1 (164) |
| **a** (Å) | 3.9335(4) | 3.9406(2) | 4.3050(2) |
| **b** (Å) | 3.9335(4) | 3.9406(2) | 4.3050(2) |
| **c** (Å) | 12.3944(17) | 12.4303(10) | 7.3157(5) |
| **V** (Å$^3$) | 191.7714(66) | 193.0218(36) | 117.4175(19) |
| Refl.collectd /unique | 725 /36 | 667 /70 | 1049 /63 |
| $R_1$ | 0.0237 | 0.0200 | 0.0209 |
| $wR_1$ | 0.0240 | 0.0257 | 0.0218 |
| Goodness-of-fit | 1.65 | 1.62 | 1.55 |

Figure S1. Microscopic pictures and compositions of different crystals grown in one batch for Sr(Fe$_{1-x}$Mn$_x$)$_2$As$_2$. (a) Phase separation with $x_0 = 0.30$. Crystal with hexagonal feature crystalizes on the plate crystal. (b) Elemental mapping of Fe and Mn on crystal at $x_0 = 0.30$. Hexagonal shaped crystal is Mn-rich, whereas the plate crystal being Fe-rich. The scale is 300 μm. (c) Miscibility gap at $x_0 = 0.36$. The crystals A and B have different morphologies and compositions. The minor Fe content in crystal B is not detectable by EDS.

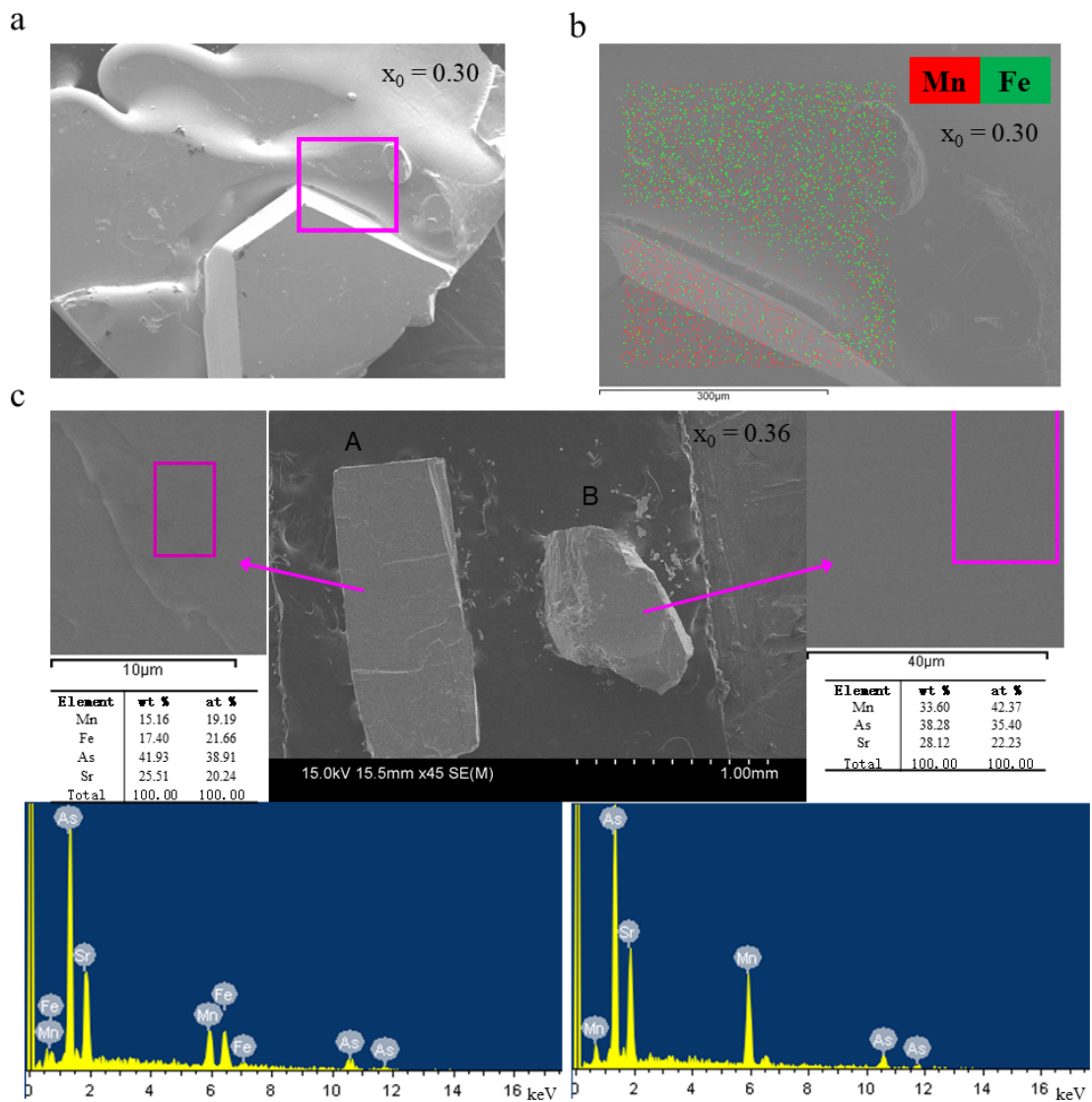

Figure S2. (Color online) The $d(\chi T)/dT$ and $dR/dT$ curves for Sr(Fe$_{1-x}$Mn$_x$)$_2$As$_2$ crystals. (a), (c) The $d(\chi T)/dT$ and $dR/dT$ curves for Sr(Fe$_{1-x}$Mn$_x$)$_2$As$_2$ crystals with $0 \leq x \leq 0.0973(1)$, respectively. (b), (d) The $d(\chi T)/dT$ and $dR/dT$ curves for Sr(Fe$_{1-x}$Mn$_x$)$_2$As$_2$ crystals with $0.0973(1) \leq x \leq 0.2055(2)$ and with $0.0973(1) \leq x \leq 0.4362(4)$, respectively. Critical temperatures were determined by the extremum of $d(\chi T)/dT$ and $dR/dT$. The arrows (Fig. S2 (a) and (c), red to black yellow)) show the suppression of SDW order, whereas the arrows (Fig. S2 (b) and (d), black yellow to blue) show the increasing temperature of broadened ordering and the arrow (Fig S2 (d), blue to purple) shows the decreasing trend of broadened ordering temperature.

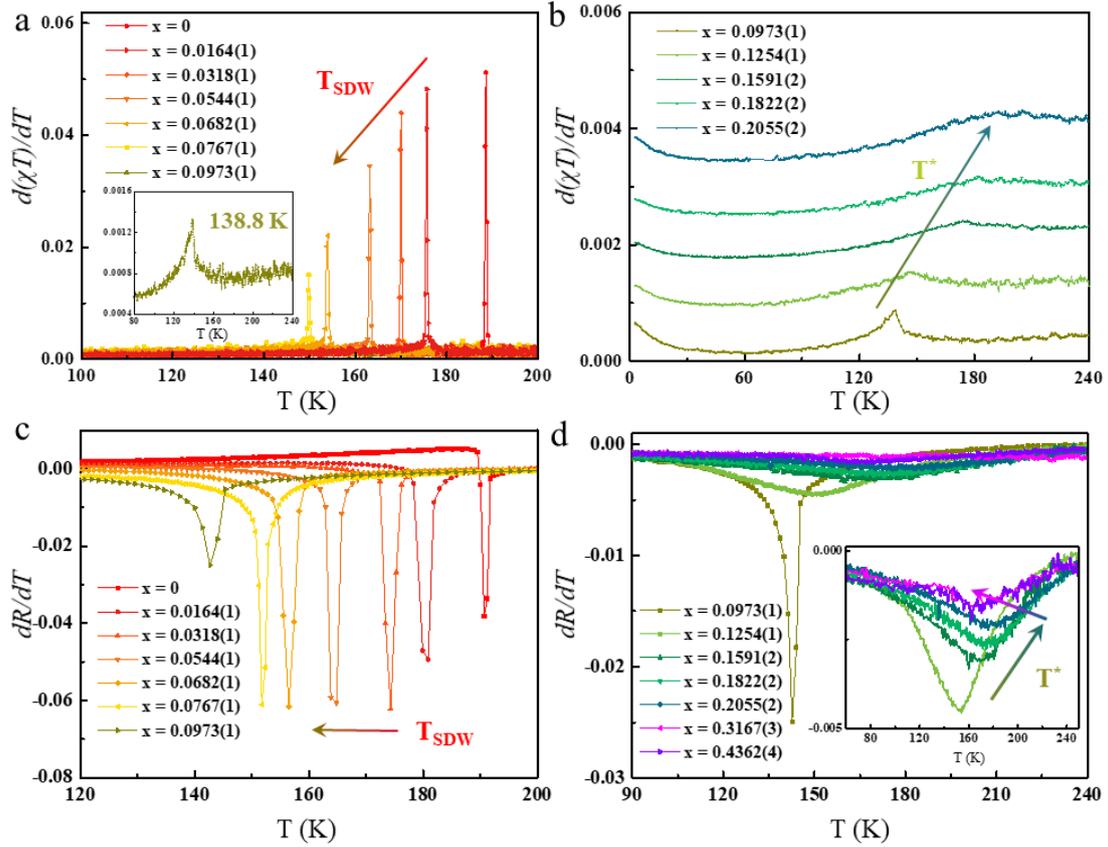

Figure S3. (Color online) Magnetic susceptibility and resistivity for Sr(Fe$_{1-x}$Mn$_x$)$_2$As$_2$ crystals with x$_0$ = 0.36. (a) The magnetic susceptibility of thin crystal with x = 0.3016(3), showing a minima around 100 K. (b) The magnetic susceptibility of thick crystal with x = 0.9612(9). The kink at 111.4 K is close to the AFM ordering temperature (125 K) of SrMn$_2$As$_2$ [3]. The deviation should be due to the Fe doping. (c) The resistivity for thin crystal with x = 0.3016(3). (d) The resistivity for thick crystal with x = 0.9612(9), which overflows at about 170 K, indicating a behavior like semiconductor. The two lines in the inset show the fitting by *ln(1/ρ) = A - Δ/T*.

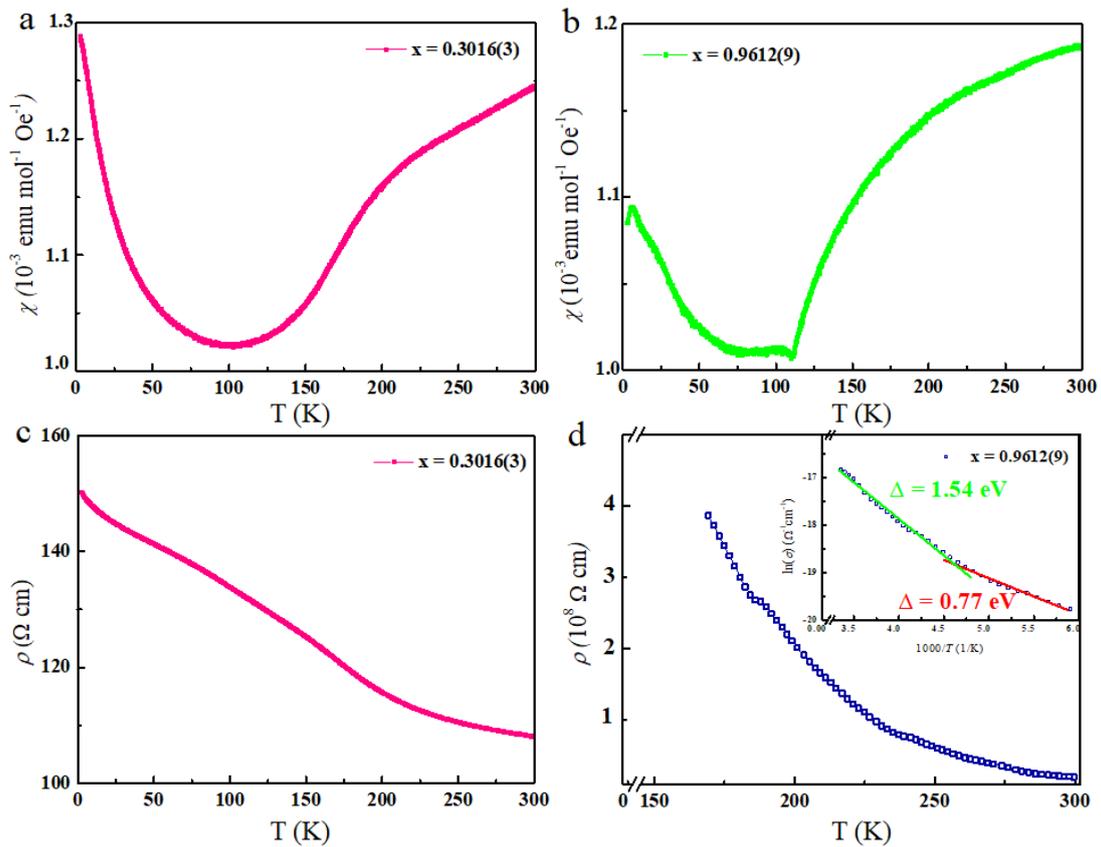